# Extrasolar 'Earths' in habitable zones: targets of opportunity

Barrie W Jones, David R Underwood, P Nick Sleep
*The Open University, Walton Hall, Milton Keynes MK7 6AA, UK*

**Abstract.** We have shown that Earth-mass planets could survive in variously restricted regions of the habitable zones (HZs) of most of a sample of nine of the 102 main-sequence exoplanetary systems confirmed by 25 September 2003. In a preliminary extrapolation of our results to the other systems, we estimate that roughly a half of these systems could have had an Earth-mass planet confined to the HZ for at least the most recent 1000 Ma. The HZ migrates outwards during the main-sequence lifetime, and so this proportion varies with stellar age – about two thirds of the systems could have such a planet confined to the HZ for at least 1000 Ma at *sometime* during the main-sequence lifetime. Clearly, these systems should be high on the target list for exploration for terrestrial planets. We have reached this conclusion by launching putative Earth-mass planets in various orbits and following their fate with mixed-variable symplectic and hybrid integrators. Whether the Earth-mass planets could *form* in the HZs of the exoplanetary systems is an urgent question that needs further study.

## 1. Introduction

It might be some years before we know for certain whether any exoplanetary systems have planets with masses the order of that of the Earth. We have therefore used a computer model to investigate a representative sample of the known exoplanetary systems, to see whether such planets could be present, and in particular whether they could remain confined to the habitable zones (HZs). If so, then it is possible that life is present on any such planets.

The HZ is that range of distances from a star where water at the surface of an Earth-like planet would be in the liquid phase. We have used boundaries for the HZ originating with Kasting, Whitmire, & Reynolds (1993). The inner boundary is the maximum distance from the star where a runaway greenhouse effect would lead to the evaporation of all surface water, and the outer boundary is the maximum distance at which a cloud-free $CO_2$ atmosphere could maintain a surface temperature of 273K. Because of simplifications in the climate model of Kasting et al, these distances are conservative in that the HZ is likely to be wider. For zero-age main-sequence stars (ZAMS stars) the HZ lies closer to the star the later its spectral type, and as the star ages the boundaries move outwards. We have revised the boundaries of Kasting et al by using a more recent model of stellar evolution (Mazzitelli 1989).

To test for confinement to the HZ of interest, we launch putative Earth-mass planets into various orbits in or near the HZ and use a mixed-variable symplectic integrator to calculate the evolution of the orbit. The integration is halted when an Earth-mass planet comes within three Hill radii ($3R_H$) of the giant planet. This is the distance at which a symplectic integrator becomes inaccurate, and is also the distance by which severe orbital perturbation of the Earth-mass planet will have occurred. Some of the integrations halted in this way were re-calculated using a hybrid integrator, where there is a switch to a Bulirsch-Stoer integrator at $3R_H$ (Chambers 1999). This enables the evolution of the orbit to be followed longer.

Confinement to the HZ over the interval of interest requires that there is no early termination of the integration by an approach within $3R_H$, *and* that the semimajor axis of the Earth-mass planet remains within the HZ. The maximum practicable pre-set integration time is constrained by the shortest orbital period (and to a lesser extent by the number of giant planets). The symplectic integrator requires a time-step shorter than about 5% of this shortest period. For systems like Rho Coronae Borealis the orbital period of the giant planet is so small that the pre-set time must be comparatively short to keep the CPU time per integration less than 100 or so hours. We used 100

Ma (100 million years) as the standard pre-set time for this system, and also for Gliese 876 and Upsilon Andromedae. For the other systems we used 1000 Ma.

Based on the Earth, 1000 Ma is the order of time required for a biosphere to begin to have an effect on a planet's surface or atmosphere that could be detected from afar, provided that we exclude any early heavy bombardment from space such as might have frustrated the emergence of life on Earth for the first 700 Ma, or possibly to several 100 Ma later (Brasier et al 2002).

## 2. The exoplanetary systems studied

We selected nine systems that between them represent a large proportion of the 102 main-sequence exoplanetary systems confirmed by 27 September 2003 (Schneider 2003). Figure 1 shows the important characteristics of these systems, ordered by increasing period of the innermost planet. The ZAMS habitable zone HZ(0) is shown shaded, and the boundaries of HZ(now) by vertical dashed lines. Each giant planet is shown by a black disc labelled with the value of $m\sin(i_0)$ in Jupiter masses $m_J$, where $m$ is the mass of the giant and $i_0$ is the inclination of the planet's orbit with respect to the plane of the sky. All but Gliese 876 have been observed only by Doppler spectroscopy, which yields $m\sin(i_0)$ rather than $m$. The motion of Gliese 876 has also been detected astrometrically, giving a value $i_0 = 84°$. At each giant planet the solid line shows the total excursion $2\Delta r$ of the giant due to its eccentricity. The dashed line extends to $(3R_H + \Delta r)$ each side of the giant when it has its minimum mass ($R_H$ is proportional to $m^{1/3}$). Further information is given in the caption. The information in Figure 1 comes largely from Schneider (2003), including the references he gives.

For an integration we have to put in an actual mass $m$ for the giant planet. This is equivalent to setting $i_0$ to some particular value. For each system we used $i_0 = 90°$, which gives the minimum value for $m$, and other values, for example $42°$, corresponding to 1.5 times the minimum.

## 3. Results

We have obtained the following results that we believe apply generally.
- The effect of the mass of the terrestrial planets has been explored from $m_{EM}$ to $8m_{EM}$, where $m_{EM}$ is the mass of the Earth plus Moon. The mass has little effect, so we restrict ourselves here to describing results with a planet of mass $m_{EM}$ and named EM.
- The presence of a second EM affects the outcome not primarily through the direct gravitational interaction between these two planets, but through close encounters between them resulting from the effect of the giant planets on each of their orbits separately. We focus on results with just one EM.
- The inclination of the orbit of EM has been explored, up to $20°$. There is little effect in most cases.
- All planets are launched with zero mean anomalies but with various longitudes of the periastra. The outcome can be sensitive to the differences $\Delta\varpi$ between these longitudes, in that some differences result in early close encounters (EM comes within $3R_H$ of the giant planet), whereas other differences result in no close encounters within the pre-set integration time. In such sensitive cases the inclination can also affect the outcome. In the case of one giant and EM we usually examined only $\Delta\varpi = 0$ and $180°$.
- At mean motion resonances with EM *interior* to the giant planet(s), instability in the orbit of EM can occur even within otherwise stable ranges of the launch value of EM's semimajor axis $a_{EM}(0)$. With EM *exterior* to the giant, the reverse holds, and stability can occur even within otherwise unstable ranges.

We have also used the nine systems to investigate the $3R_H$ criterion as a threshold of orbital instability, as outlined in the 'Introduction'. This is also the distance at which we chose to halt symplectic integrations. For EM interior to the giant planet(s) we note that symplectic integrations are typically halted by secular increases in the eccentricity of EM orbits launched near to $3R_H$ of the giant. This indicates that the $3R_H$ criterion is a useful one. The semimajor axis of EM changes little up to this point so the threshold is breached mainly by the eccentricity increase. The hybrid integrator has been used to explore the fate of some of the halted orbits within $3R_H$. In almost all cases the outcome is ejection of EM to an astrocentric distance beyond 100 AU.

For EM exterior to the giant planet the $3R_H$ criterion for halting symplectic integration again stands up. In fact the requirement for orbital stability is more severe, with secular increases in the eccentricity of EM when it is launched within about 7-8 $R_H$ of the giant.

Here is a brief summary of the integration results on each system. You can compare these brief statements with the system diagrams in Figure 1 and its caption to see that the statements and diagrams are consistent.

*Upsilon Andromedae and Gliese 876:* confinement *nowhere* in HZ(0) and HZ(now) – the HZ is traversed by $(3R_H + \Delta r)$.

*Rho Coronae Borealis*: confinement everywhere in HZ(0) and HZ(now), even at eight times the minimum giant mass.

*HD52265*: confinement almost everywhere in HZ(0) and HZ(now).

*47 Ursae Majoris:* confinement only at the innermost part of HZ(0) and HZ(now). Our conclusions are in accord with (less extensive) work done by others (Laughlin et al 2002, Noble et al 2002).

*HD196050*: confinement obtained nowhere in HZ(now), and only in the innermost part of HZ(0) when the giant planet is close to its minimum mass.

*HD216435 (Tau[1] Gruis)*: confinement was obtained only in the inner region of HZ(0). NB The parameters of this system have recently been revised, and confinement is no longer likely anywhere in HZ(0).

*HD72659*: confinement almost everywhere in HZ(0) and HZ(now).

*Epsilon Eridani*: confinement obtained nowhere in HZ(0) and HZ(now), though towards the end of the main sequence lifetime the inner HZ would provide confinement.

Further details of our work on Ups And, Gl876, Rho CrB, 47 UMa, and Eps Eri are in Jones et al (2001), Jones and Sleep (2002), Jones and Sleep (2003).

## 4. Conclusions: extension to other systems

In order to avoid the extensive integrations necessary to establish the confinement or otherwise in other exoplanetary systems, we have applied the $nR_H$ criteria at the HZ boundaries in each of these systems.

Figure 2 shows a notional HZ migrating outwards during the main sequence. Consider first a single giant planet closer to the star than the HZ, and suppose that it has the apoastron and periastron distances shown, with $7R_H$ extending outwards from apoastron. The whole HZ lies beyond $7R_H$ for the whole of the main-sequence, so we conclude that confined EM orbits are likely anywhere in the HZ at any time during the main-sequence. Of particular interest from a biological perspective is whether there could have been such orbits throughout at least the most recent 1000 Ma, excluding the first 700 Ma of the main sequence, as discussed in the 'Introduction'. It is also of interest whether such a span of 1000 Ma can be found at any time during the main-sequence, and not just recently.

Figure 2 also shows a single giant further from the star, with $3R_H$ extending inwards from periastron. The whole HZ lies beyond $3R_H$ for the whole of the main-sequence, so we again conclude that confined EM orbits are likely anywhere in the HZ at any time during the main-sequence. In this case too the most recent 1000 Ma and any span of 1000 Ma are of biological interest.

Table 1 summarises the results of this kind of analysis to all of the main-sequence exoplanetary systems in Schneider (2003). The following conventions are adopted.
1  The systems are in the order listed by Schneider, by increasing period of the planet with the shortest period. The correct sequence is *down* the table columns, not across.

2  The $nR_H$ are calculated using the minimum giant masses, though $R_H$ varies slowly, as $m^{1/3}$.
3  Systems with more than one planet are shown italicised.
4  The column 'now' shows whether an EM could be confined to the HZ within at least the past 1000 Ma (excluding the first 700 Ma of the main-sequence). If the entry is 'yes' then it could do so almost anywhere in the HZ. If the entry is 'NO' then nowhere in the HZ should offer confinement. If the entry is 'part' then some small proportion of the HZ should offer confinement, for example near its outer boundary for a giant planet not much closer to the star than the inner boundary.
5  The column 'sometime' refers to whether an EM could be confined to the HZ for at least 1000 Ma at any time in the main-sequence (again excluding the first 700 Ma).
6  A '?' denotes a star of unknown age, where this is crucial to the evaluation.
7  A '**' denotes those very few cases where the periastron of the giant lies beyond the HZ even at the end of the main-sequence. These are the systems most like the Solar System.

With Table 1 ordered by increasing period, the entries start with hot-Jupiters. These are well interior to the HZ throughout the main-sequence, and so confinement is likely across the whole HZ at all times. Exceptions are among the multiple planet systems (italicised in Table 1). In these, the outer giant(s) compromise(s) confinement. We then move to 'warm-Jupiters', around HD3651, and confinement is compromised even when there is just a single giant planet. Around GJ3021 the giant orbits in or near the inner part of HZ(0). The trend down the table is then for the giant(s) to move across the HZ. There are as yet only four exoplanetary systems where the periastron of the giant lies beyond the HZ throughout the main-sequence.

Overall, we estimate that roughly a half of the systems in Table 1 could have had an Earth-mass planet confined to the HZ for at least the most recent 1000 Ma ('now' = 'yes' or 'part'), and that about two thirds of the systems could have such a planet confined to the HZ for at least a billion years *sometime* during the main-sequence lifetime ('sometime' = 'yes' or 'part').

Whether the Earth-mass planets could form in the HZs of the exoplanetary systems is an urgent question that needs further study.

## References

Brasier, M.D., et. al, 2002, Nature, 416, 76
Chambers, J.E. 1999, MNRAS, 305, 793
Jones, B.W., Sleep, P.N., & Chambers, J.E. 2001, A&A, 366, 254
Jones, B.W., & Sleep, P.N. 2002, A&A, 393, 1015
Jones, B.W., & Sleep, P.N. 2003, ASP Conf Ser, 294, 225
Kasting, J.F., Whitmire D.P., & Reynolds, R.T. 1993, Icarus, 101, 108
Laughlin, G., Chambers, J.E., & Fischer, D.A. 2002, ApJ, 579, 455
Mazzitelli, I. 1989, ApJ, 340, 249
Noble, M., Musielak, Z.E., & Cuntz, M. 2002, ApJ, 572, 1024
Schneider, J. 2003, http://www.obspm.fr/encycl/cat1.html

Table 1  Habitability of extrasolar planetary systems.

| Star | now | sometime | Star | now | sometime | Star | now | sometime |
|---|---|---|---|---|---|---|---|---|
| OG-TR-56 | yes | yes | HD80606 | NO | part | *HD160691* | *NO* | *NO* |
| HD73256 | yes | yes | HD219542B | yes | yes | HD141937 | NO | NO |
| HD83443 | yes | yes | 70 Vir | ? | yes | HD41004A | part | part |
| HD46375 | yes | yes | HD216770 | ? | yes | HD47536 | NO | NO |
| HD179949 | yes | yes | HD52265 | yes | yes | HD23079 | NO | part |
| HD187123 | yes | yes | GJ3021 | NO | yes | 16 CygB | NO | NO |
| Tau Boo | yes | yes | *HD37124* | *NO* | *NO* | HD4208 | NO | NO |
| BD103166 | yes | yes | HD73526 | part | yes | HD114386 | yes** | yes** |
| HD75289 | yes | yes | HD104985 | yes | yes | Gam Ceph | part | part |
| HD209458 | yes | yes | *HD82943* | *NO* | *part* | HD213240 | NO | NO |
| HD76700 | yes | yes | *HD169830* | *part* | *part* | HD10647 | NO | part |
| 51 Peg | yes | yes | HD8574 | part | yes | *HD10697* | *NO* | *NO* |
| *Ups And* | *NO* | *NO* | HD89744 | part | yes | *47 UMa* | *part* | *part* |
| HD49674 | yes | yes | HD134987 | part | yes | HD190228 | NO | NO |
| HD68988 | yes | yes | HD40979 | part | yes | HD114729 | NO | NO |
| HD168746 | yes | yes | *HD12661* | *NO* | *NO* | HD111232 | ? | part |
| HD217107 | yes | yes | HD150706 | ? | yes | HD2039 | NO | NO |
| HD162020 | yes | yes | HR810 | part | yes | HD136118 | NO | NO |
| HD130322 | yes | yes | HD142 | ? | yes | HD50554 | NO | NO |
| HD108147 | yes | yes | HD92788 | NO | part | HD196050 | NO | part |
| *HD38529* | *NO* | *NO* | HD28185 | NO | part | HD216437 | ? | part |
| *55 Cancri* | *yes* | *yes* | HD142415 | ? | part | HD216435 | NO | NO |
| Gliese 86 | yes | yes | HD177830 | part | yes | HD106252 | NO | NO |
| HD195019 | yes | yes | HD108874 | NO | part | HD23596 | NO | NO |
| HD6434 | yes | yes | HD4203 | part | part | 14 Her | NO | NO |
| HD192263 | yes | yes | HD128311 | NO | NO | HD39091 | ? | part |
| *Gliese 876* | *NO* | *NO* | HD27442 | part | yes | HD72659 | yes** | yes** |
| Rho CrB | yes | yes | HD210277 | part | part | HD70642 | yes** | yes** |
| *HD74156* | *NO* | *NO* | HD19994 | yes | yes | HD33636 | NO | NO |
| HD168443 | NO | NO | HD20367 | ? | yes | *Eps Eridani* | *NO* | *part* |
| HD3651 | ? | yes | HD114783 | NO | part | HD30177 | part | part |
| HD121504 | yes | yes | HD147513 | NO | NO | Gliese 777A | yes** | yes** |
| HD178911B | ? | yes | HIP75458 | NO | NO | | | |
| HD16141 | yes | yes | HD222582 | NO | NO | | | |
| HD114762 | NO | part | HD65216 | ? | part | | | |

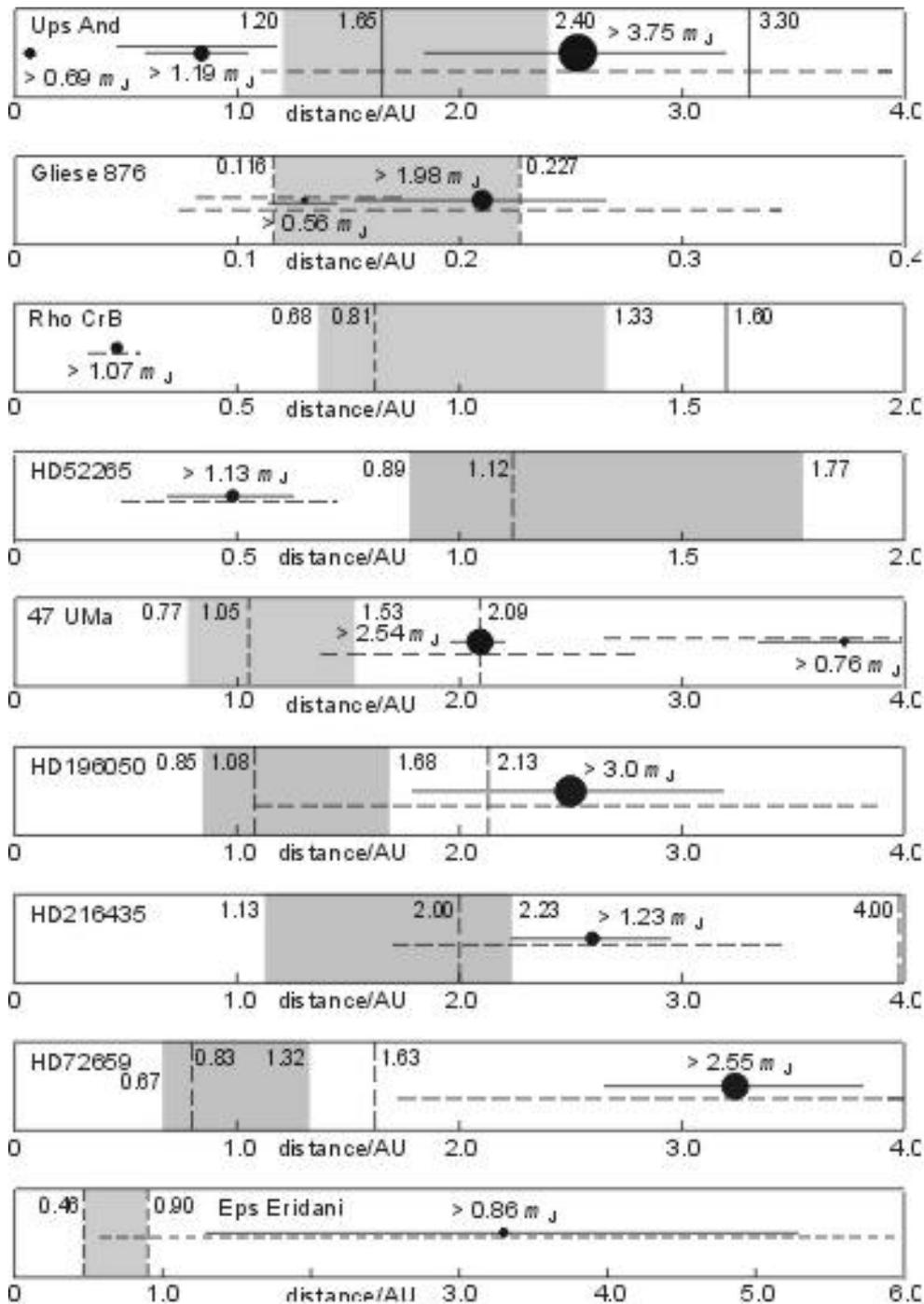

Figure 1. The nine exoplanetary systems studied, ordered by increasing period of the innermost planet. At each giant planet (solid disc) the solid line shows the total excursion $2\Delta r$ of the giant due to its eccentricity. The dashed line extends to $(3R_H + \Delta r)$ each side of the giant when it has its minimum mass. The stars are all main sequence, with masses (in solar masses), [Fe/H], and ages (in Ma) as follows. Ups And 1.3, 0.09, 3300. Gl876 0.32, 0, (unknown). Rho CrB 0.95, -0.19, 6000. HD52265: 1.13, 0.11, 4500. 47 UMa 1.03, -0.08, 7000. HD196050 1.1, 0-0.25, 5000. HD216435 1.25, 0.15, 5000. HD72659 0.95, -0.14, 7000. Eps Eri 0.8, -0.1, 500-1000.

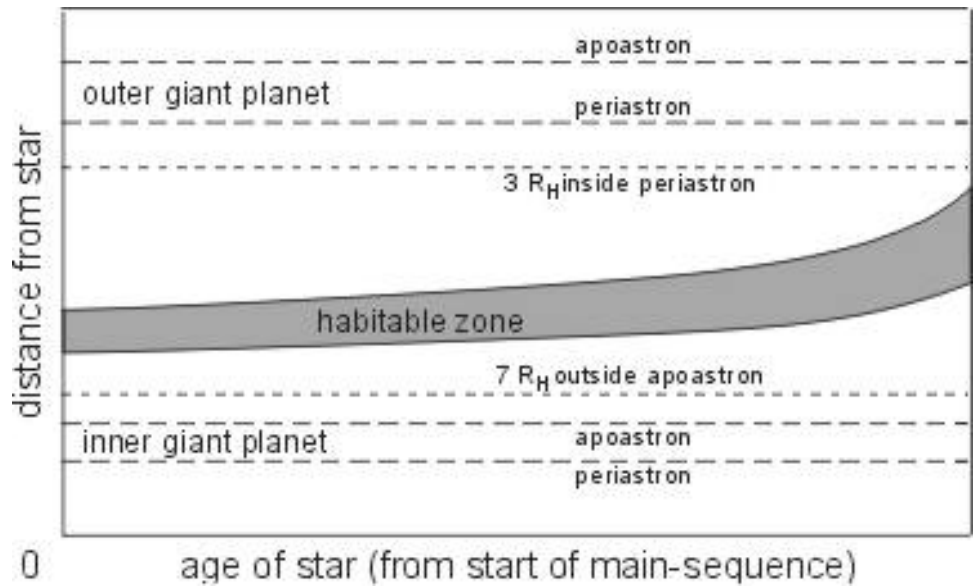

Figure 2. A notional habitable zone (HZ) migrating outwards during the main-sequence, with a giant planet interior to the HZ, and, alternatively, a giant planet exterior to the HZ. $R_H$ is the Hill radius of the giant planet.